\documentclass[twoside,twocolumn,english]{revtex4-1}
\usepackage{mathptmx}
\usepackage[T1]{fontenc}
\usepackage[latin9]{inputenc}
\usepackage{verbatim}
\usepackage{float}
\usepackage{units}
\usepackage{amsmath}
\usepackage{graphicx}
\usepackage[colorlinks = true,
            linkcolor = blue,
            urlcolor  = blue,
            citecolor = blue,
            anchorcolor = blue]{hyperref}
\usepackage{epstopdf}

\makeatletter

 \@ifundefined{textcolor}{}
 {%
   \definecolor{BLACK}{gray}{0}
   \definecolor{WHITE}{gray}{1}
   \definecolor{RED}{rgb}{1,0,0}
   \definecolor{GREEN}{rgb}{0,1,0}
   \definecolor{BLUE}{rgb}{0,0,1}
   \definecolor{CYAN}{cmyk}{1,0,0,0}
   \definecolor{MAGENTA}{cmyk}{0,1,0,0}
   \definecolor{YELLOW}{cmyk}{0,0,1,0}
 }

\makeatother

\begin{document}

\title{Heralded generation of Bell states using atomic ensembles}

\author{David Shwa, Raphael D. Cohen, Alex Retzker and Nadav Katz }



\affiliation{Racah Institute of Physics, The Hebrew University of Jerusalem, Jerusalem 91904, Israel }
\begin{abstract}
We propose a scheme that utilizes the collective enhancement of a
photonic mode inside an atomic ensemble together with a proper Zeeman
manifold in order to achieve a heralded polarization entangled Bell
state. The entanglement is between two photons that are separated
in time and can be used as a post selected deterministic source for
applications such as quantum repeaters where a subsequent entanglement
swapping measurement is employed. We present a detailed analysis of
the practical limitation of the scheme.
\end{abstract}
\maketitle

\section{Introduction}

Entanglement is a unique property of quantum multisystems, where the
state of one system is not independent on the others \cite{Pan2012}.
Entanglement serves as the main tool in fundamental research of quantum
theory as well as in the rapidly-developing area of quantum information
\cite{Horodecki2009}. Photons are prominent quantum systems due to their
weak interaction with the environment which increase their immunity
to decoherence. On the other side this weak interaction makes the creation
of an entangled state of photons a difficult task that usually requires
a very high nonlinearity. In the early days of quantum optics sources
of entanglement were atomic cascades \cite{Aspect1981}, but nowadays
the main source for entangled photons is the nonlinear process of spontaneous
parametric down conversion (SPDC). This is an efficient source that can create
polarization \cite{Kwiat1995} or time-bin entanglement \cite{Kwiat1993},
but has two major drawbacks for efficient quantum communication schemes.
Namely, it is not deterministic and has a broadband spectrum. Deterministic
single photon sources include quantum dots, single atoms in a cavity
and atomic ensembles \cite{Eisaman2011,Kuhn2002}. Atomic ensembles
offer another asset, which is the generation and storage of a single
photon in a heralded way. This is the main building block for a quantum
repeater as proposed in the DLCZ protocol for long range quantum communication
\cite{Duan2001,Sangouard2009}. Single photon storage times of up
to a few ms were observed using trapped Rubidium ensembles \cite{Zhao2008a,Zhao2008b}
and a few tens of $\mu s$ using warm vapor \cite{VanderWal2003,Bashkansky2012}.
Moreover, the ability to store a multi photon entangled state from
an SPDC source was also shown \cite{Dai2012,Choi2008}. Recently,
several alternatives to SPDC as an entanglement source were presented.
Quantum dot biexcitons were developed as an efficient source for entangled
photons that can be created in a triggered way \cite{Akopian2006,Stevenson2006}.
Using a single quantum dot ensures a single pair of entangled photons,
but the yield up until now is not high compared to SPDC. Moreover
the photons are emitted together and are still broadband with respect
to the needs of quantum repeaters \cite{Sangouard2009}. New schemes
of exploiting atomic media as an entanglement source were also presented.
One proposal uses a double-$\Lambda$ level configuration for a deterministic
entanglement of N photons \cite{Gheri1998}. This procedure suffers
mostly from the difficulty of working with one atom in a cavity. Another
promising direction for entangled photon source is the use of non linear effects in atomic ensembles such as four wave mixing \cite{Srivathsan2013,MacRae2012} and Rydberg
blockade \cite{Porras2008,Nielsen2010}.    
Porras and Cirac suggested a use of an excited symmetric spin wave
in double $\Lambda$ atoms as a way to entangle photons in a deterministic
way \cite{Porras2008}.

Here we take this idea in a different direction and apply it in an atomic ensemble. We utilize single photon quantum
storage in atomic gases combined with the property of Zeeman splitting
of hyperfine manifolds in order to create a heralded polarization
entanglement. The scheme relies on using the magnetic Zeeman levels
as an effective polarization beam splitter for single photons in order
to entangle the two photons. This source creates two polarization
entangled photons that are distinguished in time and have a narrow
bandwidth that can be suitable for quantum communication \cite{WALTHER2007}.
We show the dependence of the fidelity and pair production rate upon
the detection efficiency. This paper is arranged as follows: in section
\ref{sec:General-Scheme} the general scheme of the entanglement process
is described. Section \ref{sec:Fidelity-and-experimental} discusses
the practical limitations of the scheme and how the fidelity and production
rate of the entangled pair is affected by them. Section \ref{sec:Conclusions}
gives some concluding remarks.

\section{General Scheme\label{sec:General-Scheme}}

\begin{figure}
\includegraphics[width=8.6cm]{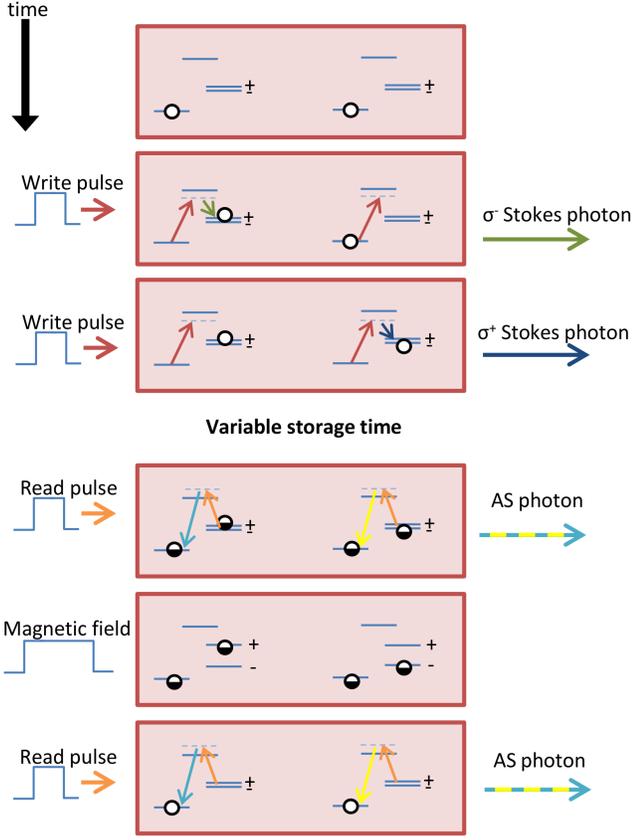}

\caption{\label{fig:schematics} Schematics of the experiment. Each of the two atoms shown represents one of the N states in the sum of the Dicke state.}  
\end{figure}

\begin{figure}
\includegraphics[width=8.6cm]{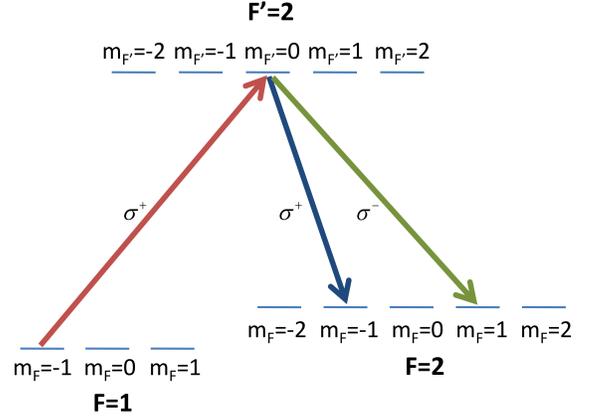}

\caption{\label{fig:write level scheme}The possible alternatives for single
Stokes photon generation during the write process. Red arrow - write
beam, Green and blue arrows - Stokes photons. }
\end{figure}

The general sequence for creating heralded entanglement is presented
schematically in Fig. \ref{fig:schematics}. As the source for entanglement
we use an atomic ensemble with N atoms. Each atom has a $\Lambda$
configuration energy level scheme. Each energy level should contain
its own Zeeman sublevel manifold, such as a hyperfine splitting with
F>0. Without loss of generality we will concentrate here on the case
where the long lived ground state $\left|g\right\rangle $ has a hyperfine
level with F=1, the long lived metastable level, $\left|s\right\rangle $
has a hyperfine level with F=2 and the excited state $\left|e\right\rangle $
is F'=2. One specific example that fits to this case is the D1 transition
of   $\:^{87}\textup{Rb}$. Each hyperfine level has Zeeman sublevels that become
non degenerate when applying a magnetic field. A schematic picture
of the relevant levels is depicted in figure \ref{fig:write level scheme}.
Using a circular polarized pumping it is possible to transfer all
the population to the $\textup{F=1},m_{s}=-1$ level which is the $\left|g^{-}\right\rangle $
state. The collective state of the atoms and light can be written
as follows :

\begin{equation}
\left|\Psi_{0}\right\rangle =\left|g^{-}\right\rangle^{N} \left|0\right\rangle, \label{eq:1}
\end{equation}
where $\left|0\right\rangle $ is the state of zero photons in a defined
spatial and spectral mode. This mode will be defined later as the
Stokes or anti Stokes (AS) mode.

Immediately after the pumping, a weak and short write pulse with a
circular $\sigma^{+}$ polarization is applied to the ensemble. The
laser detuning should be large enough in order for the main atom light
interaction to be a spontaneous Raman transition to the $\left|s\right\rangle $
state. Since the circular polarization dictates a transition to the
Zeeman sublevel $\left|e,m_{s}=0\right\rangle $, the spontaneous
Raman decay can be to levels $\left|s,m_{s}=0,\pm1\right\rangle $,
but since $\textup{F}=\textup{F}'$ the transition to $\left|s,m_{s}=0\right\rangle $
is forbidden and the other two sublevels have the same probability
\cite{circular}. Thus, upon a successful detection of one photon
in one of the polarizations $\sigma^{+}$ or $\sigma^{-}$, the spin
state of the metastable level will become $\left|s,m_{s}=-1\right\rangle $
or $\left|s,m_{s}=+1\right\rangle $ respectively. If for example
the detected photon is $\sigma^{+}$ then the collective state will
be 

\begin{equation}
\left|\Psi_{s1}\right\rangle =\frac{1}{\sqrt{N}}\sum_{i=1}^{N}e^{i\mathbf{(k_{w}-k_{s})x_{i}}}\left|s_{i}^{-}\right\rangle \left|+_{s}\right\rangle, \label{eq:2}
\end{equation}
where $\left|s_{i}^{-}\right\rangle $ means that the $i$th atom
is now at the state $\left|s,m_{s}=-1\right\rangle $ while all other atoms are in the ground $\left|g^{-}\right\rangle $ state.  $\left|+_{s}\right\rangle $
means a Stokes photon with $\sigma^{+}$ polarization is emitted,
$\mathbf{k_{w}}$ is the write laser wave vector, $\mathbf{k_{s}}$
is the Stokes photon wave vector and $\mathbf{x_{i}}$ is the coordinate
of the $i$th atom. The atomic state is in a Dicke-like state
\cite{Dicke1954}. This is a long lived atomic coherence where the main decoherence
process is the atomic motion that can change the phases between the
different atomic states \cite{Duan2002}. In the following, we will
assume no motion, meaning that these phases do not alter during the
storage time. In this case, upon applying a read pulse with a perfect
phase matching condition $\mathbf{k_{w}}+\mathbf{k_{r}=\mathbf{k_{s}}+\mathbf{k_{as}}}$,
a constructive interference in the direction of the AS wave
vector will cause this mode to dominate all other directions \cite{Sangouard2009}.
Thus, in the case of no atomic motion and perfect phase matching,
the phase terms just sum up to unity so it is possible to omit them from
now on. 

Now it is possible to quickly repeat the write pulse in order to create
another excitation in the metastable state (of course, this pulse
will create an excitation only with low probability, but since it
is a heralded scheme, we are looking only upon successful events).
Let us concentrate on the events where the second Stokes photon that
is emitted has the opposite polarization than the first one, thus
the collective state will be now

\begin{equation}
\left|\Psi_{s2}\right\rangle =\frac{1}{\sqrt{2N(N-1)}}\sum_{i\neq j=1}^{N}\left|s_{i}^{-}s_{j}^{+}\right\rangle \left|+_{s}-_{s}\right\rangle. \label{eq:3}
\end{equation}

\begin{figure}
\includegraphics[width=8.6cm]{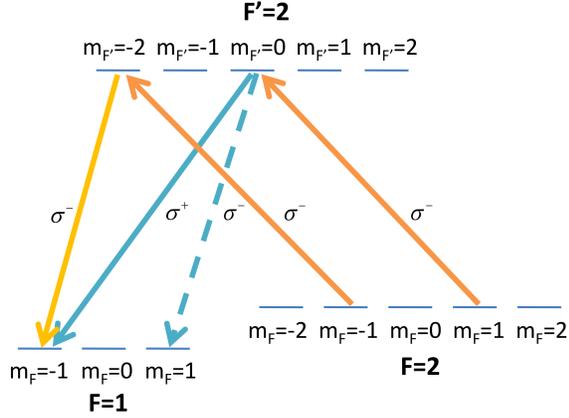}

\caption{\label{fig:Anti-stokes level scheme}The possible alternatives for
single AS photon generation during the read process. orange arrows - read beam, blue, dashed blue and yellow arrows -
AS photons.}
\end{figure}

After a certain storage time, a $\sigma^{-}$
polarization read pulse is sent into the media. This  pulse can interact with one of the excited
Dicke states releasing with some probability one AS photon. There are two possibilities, one that the atoms are in the $\left|s,m_{s}=+1\right\rangle $
state and one that the atoms are in the $\left|s,m_{s}=-1\right\rangle $
state. Each of these options can produce different AS single photons
as described in Fig. \ref{fig:Anti-stokes level scheme}. The AS
transition strengths may not be the same due to different Clebsch\textendash{}Gordan
coefficients, thus there is a need to multiply each transition with
the proper probability amplitude denoted by $P_{kl}^{(\textup{F})}$ where $k$ is the
Zeeman sublevel of the excited state, $l$ is the Zeeman sublevel
of the ground state and $\textup{F}$ is the hyperfine level of the ground state.

The AS photon polarization is correlated with the atomic spin level
that remains in the ensemble and the state becomes
\begin{equation}
{\displaystyle \begin{array}{c}
\left|\Psi_{as1}^{0}\right\rangle =\frac{1}{\sqrt{2N(N-1)}}{\displaystyle \sum_{i\neq j=1}^{N}}{\scriptstyle P}_{ {\scriptscriptstyle 0,1}}^{ {\scriptscriptstyle (2)}}\left|s_{i}^{-}\right\rangle ({\scriptstyle P}_{{\scriptstyle {\scriptscriptstyle 0,-1}}}^{ {\scriptscriptstyle (1)}}\left|\text{g}_{j}^{-}\right\rangle \left|+_{as}\right\rangle \\+{\scriptstyle P}_{{\scriptscriptstyle 0,1}}^{ {\scriptscriptstyle (1)}}\left|\text{g}_{j}^{+}\right\rangle \left|-_{as}\right\rangle )
+{\scriptstyle P}_{{\scriptscriptstyle -2,-1}}^{ {\scriptscriptstyle (2)}}{\scriptstyle P}_{{\scriptscriptstyle -2,-1}}^{ {\scriptscriptstyle (1)}}\left|s_{i}^{+}\right\rangle \left|\text{g}_{j}^{-}\right\rangle \left|-_{as}\right\rangle,
\end{array}}\label{eq:4}
\end{equation}
where $\left|\text{g}^{-}\right\rangle $ and $\left|\text{g}^{+}\right\rangle $
are the relaxation of the first AS photon to the Zeeman ground state
sublevels $m_{F}=-1$ or $m_{F}=+1$ respectively.

Now a magnetic field is applied to the ensemble creating a Zeeman
splitting of the $\left|s\right\rangle $, $\left|g\right\rangle $
states. The two Zeeman levels will acquire a different phase during
the single excitation storage time according to the energy splitting.
For an energy splitting $\omega_{m}$ in the $\left|s\right\rangle $
level and $\omega_{n}$ in the $\left|g\right\rangle $ level and
after a storage time $\tau$ the state will become

\begin{equation}
  {\textstyle {\scriptstyle \begin{array}{l}
{\displaystyle \left|\Psi_{as1}^{\tau}\right\rangle ={\scriptstyle {\textstyle \frac{1}{\sqrt{2N(N-1)}}}}\sum_{i\neq j=1}^{N}{\scriptstyle P}_{{\scriptscriptstyle 0,1}}^{ {\scriptscriptstyle (2)}}e^{-i\omega_{m}\tau}\left|s_{i}^{-}\right\rangle}\\\;\;\;\;\times{\displaystyle ({\scriptstyle P}_{{\scriptscriptstyle 0,-1}}^{ {\scriptscriptstyle (1)}}e^{-i\omega_{n}\tau}\left|\text{g}_{j}^{-}\right\rangle \left|+_{as}\right\rangle 
 + {\scriptstyle P}_{{\scriptscriptstyle 0,1}}^{ {\scriptscriptstyle (1)}}e^{i\omega_{n}\tau}\left|\text{g}_{j}^{+}\right\rangle \left|-_{as}\right\rangle )}\\
\;\;\;\;{\displaystyle +{\scriptstyle P}_{{\scriptscriptstyle -2,-1}}^{ {\scriptscriptstyle (2)}}{\scriptstyle P}_{{\scriptscriptstyle -2,-1}}^{ {\scriptscriptstyle (1)}}e^{i(\omega_{m}-\omega_{n})\tau}\left|s_{i}^{+}\right\rangle \left|\text{g}_{j}^{-}\right\rangle \left|-_{as}\right\rangle .}
\end{array}}}
\end{equation}

Sending another read pulse strong enough to create a second AS photon will produce the following state \cite{unity}:
\\

\begin{widetext}

\begin{equation}
\begin{array}{c}
{\textstyle {\displaystyle \left|\Psi_{as2}^{\tau}\right\rangle ={\displaystyle {\textstyle \frac{\alpha}{\sqrt{2N(N-1)}}}\sum_{i\neq j=1}^{N}}e^{-i\omega_{m}\tau}{\textstyle \left|\text{g}_{i}^{-}\right\rangle ^{{\scriptscriptstyle B}}}\left|-_{as}\right\rangle ^{{\scriptscriptstyle B}}({\scriptstyle P}_{{\scriptscriptstyle 0,1}}^{ {\scriptscriptstyle (1)}}e^{i\omega_{n}\tau}\left|\text{g}_{j}^{+}\right\rangle ^{{\scriptscriptstyle A}}\left|-_{as}\right\rangle ^{{\scriptscriptstyle A}}}}
+{\scriptstyle P}_{{\scriptscriptstyle 0,-1}}^{ {\scriptscriptstyle (1)}}e^{-i\omega_{n}\tau}\left|\text{g}_{j}^{-}\right\rangle ^{{\scriptscriptstyle A}}\left|+_{as}\right\rangle ^{{\scriptscriptstyle A}})+e^{i(\omega_{m}-\omega_{n})\tau}({\scriptstyle P}_{{\scriptscriptstyle 0,-1}}^{ {\scriptscriptstyle (1)}}\left|\text{g}_{i}^{-}\right\rangle ^{{\scriptscriptstyle B}}\left|+_{as}\right\rangle ^{{\scriptscriptstyle B}}\\
+{\scriptstyle P}_{{\scriptscriptstyle 0,1}}^{ {\scriptscriptstyle (1)}}\left|\text{g}_{i}^{+}\right\rangle ^{{\scriptscriptstyle B}}\left|-_{as}\right\rangle ^{{\scriptscriptstyle B}})\left|\text{g}_{j}^{-}\right\rangle ^{{\scriptscriptstyle A}}\left|-_{as}\right\rangle ^{{\scriptscriptstyle A}},
\end{array}\label{eq:6}
\end{equation}

\end{widetext}
where the $A,B$ notations represent the emitted AS photon during
the first or second read pulse respectively and $\alpha={\scriptstyle P}_{{\scriptscriptstyle 0,1}}^{ {\scriptscriptstyle (2)}}{\scriptstyle P}_{{\scriptscriptstyle -2,-1}}^{ {\scriptscriptstyle (2)}}{\scriptstyle P}_{{\scriptscriptstyle -2,-1}}^{ {\scriptscriptstyle (1)}}$. For simplicity in the
following we will abbreviate $\left|-_{as}\right\rangle ^{A}\left|-_{as}\right\rangle ^{B}\equiv\left|--\right\rangle $
etc., hence the state can be written as 

\begin{equation}
\begin{array}{c}
\left|\Psi_{as2}^{\tau}\right\rangle ={\textstyle \frac{\alpha}{\sqrt{2N(N-1)}}{\textstyle {\displaystyle \sum_{i\neq j=1}^{N}}}}e^{-i(\omega_{m}+\omega_{n})\tau}{\scriptstyle P}_{{\scriptscriptstyle 0,-1}}^{ {\scriptscriptstyle (1)}}\left|\text{g}_{j}^{-}\text{g}_{i}^{-}\right\rangle \left|+-\right\rangle \\
+e^{-i(\omega_{m}-\omega_{n})\tau}{\scriptstyle P}_{{\scriptscriptstyle 0,1}}^{ {\scriptscriptstyle (1)}}\left|\text{g}_{j}^{+}\text{g}_{i}^{-}\right\rangle \left|--\right\rangle \\
+e^{i(\omega_{m}-\omega_{n})\tau}{\scriptstyle P}_{{\scriptscriptstyle 0,-1}}^{ {\scriptscriptstyle (1)}}\left|\text{g}_{j}^{-}\text{g}_{i}^{-}\right\rangle \left|-+\right\rangle \\
+e^{i(\omega_{m}-\omega_{n})\tau}{\scriptstyle P}_{{\scriptscriptstyle 0,1}}^{ {\scriptscriptstyle (1)}}\left|\text{g}_{j}^{-}\text{g}_{i}^{+}\right\rangle \left|--\right\rangle.
\end{array}\label{eq:7}
\end{equation}

Since there is a sum over all the ensemble, the time ordering may
be switched and different indices for the atoms can be dropped. As
$\left|\text{g}^{-}\right\rangle $ is just the ground state it can
be omitted from the equation and we get 

\begin{equation}
\begin{array}{c}
\left|\Psi_{as2}^{\tau}\right\rangle =2\alpha{\scriptstyle P}_{{\scriptscriptstyle 0,1}}^{ {\scriptscriptstyle (1)}}\cos[(\omega_{m}-\omega_{n})\tau]\left|--\right\rangle {\textstyle {\displaystyle (\frac{1}{N}\sum_{i=1}^{N}}}\left|\text{g}_{i}^{+}\right\rangle )\\
+\alpha{\scriptstyle P}_{{\scriptscriptstyle 0,-1}}^{ {\scriptscriptstyle (1)}}(e^{i(\omega_{m}-\omega_{n})\tau}\left|-+\right\rangle +e^{-i(\omega_{m}+\omega_{n})\tau}\left|+-\right\rangle ).
\end{array}
\end{equation}

Let us normalize the state for every storage time, thus the normalized
state will be (taking into account that $P_{{\scriptscriptstyle 0,1}}^{ {\scriptscriptstyle (1)}}=P_{{\scriptscriptstyle 0,-1}}^{ {\scriptscriptstyle (1)}}$)

\begin{equation}
\begin{array}{l}
\left|\Psi_{as2}^{\tau}\right\rangle =\sqrt{2}\frac{\cos[(\omega_{m}-\omega_{n})\tau]}{\sqrt{2\cos^{2}[(\omega_{m}-\omega_{n})\tau]+1}}\left|--\right\rangle {\textstyle ({\displaystyle \frac{1}{N}\sum_{i=1}^{N}}}\left|\text{g}_{i}^{+}\right\rangle )\\
+\frac{1}{\sqrt{2}}\frac{1}{\sqrt{2\cos^{2}[(\omega_{m}-\omega_{n})\tau]+1}}\left(e^{i(\omega_{m}-\omega_{n})\tau}\left|-+\right\rangle +e^{-i(\omega_{m}+\omega_{n})\tau}\left|+-\right\rangle \right).
\end{array}\label{eq:8}
\end{equation}

The second term is actually a generalized Bell state. The most interesting case will be for storage time where the phase is $(\omega_{m}-\omega_{n})\tau=\unitfrac{\pi}{2}$.
In this case the first term vanishes and the photonic state is just
a rotated maximally entangled Bell state $\left|\Psi'\right\rangle =\frac{1}{\sqrt{2}}\left(\left|-+\right\rangle +e^{i\phi}\left|+-\right\rangle \right)$.

\section{Practical issues\label{sec:Fidelity-and-experimental}}

The previous section dealt with an ideal case. For real applications
there are a few issues that needs to be answered regarding the scheme
we presented. The main two issues are the coherence time and the fidelity
of the process.

\subsection{Coherence}

In order for the scheme to succeed the coherence time
of the collective state of the atoms should be much longer than the
experiment time. A typical coherence can reach up to 1 ms in cold
atoms and a few hundred $\mu s$ in warm vapor \cite{Zhao2008a}.
In warm vapor the spatial coherence of the collective spin state limits
the lifetime. Measurements in Rubidium of single photon storage reveal
a quantum nature up to 5 $\mu s$ \cite{Bashkansky2012}.
For a 0.1 $\mu s$ time storage with magnetic field, the frequency shift
for substantial phase shift will be on the order of 10 MHz meaning
a magnetic field of $\sim$10 G. Switching on and off such
a field with a 100 ns time scale is achievable.

\subsection{Fidelity}

The fidelity of the entangled state is affected by three major contributions.
The first one is having more than one excited Stokes/anti Stokes photon
per write/read pulse, the second one is the detection efficiency and
the third one is the detector's dark counts \cite{Sangouard2009}.
In general, for spontaneous Raman scattering the state after a write
pulse can be written as

\begin{equation}
\begin{array}{c}
\left|\Psi\right\rangle =\sqrt{P_{\lambda}(0)}\left|e\right\rangle \left|0_{s}\right\rangle +\sqrt{P_{\lambda}(1)}\frac{1}{\sqrt{N}}\sum_{i=1}^{N}\left|s_{i}\right\rangle \left|1_{s}\right\rangle \\
+\sqrt{P_{\lambda}(2)}\frac{1}{N}\sum_{i=1}^{N}\sum_{j=1}^{N}\left|s_{i}s_{j}\right\rangle \left|2_{s}\right\rangle +....,
\end{array}
\end{equation}
where $P_{\lambda}(n)$ is the probability of exciting $n$ atoms in
the specific mode and $\left|n_{s}\right\rangle $ is a state with
$n$ Stokes photons. For low excitation number each excitation is independent,
thus this probability will have a Poisson distribution defined by
parameter $\lambda$, which is the average excitation number. In order
to have only a single excitation, at most, due to each write pulse we need to use
a weak pulse such that $P_{\lambda}(0)\gg P_{\lambda}(1)\gg P_{\lambda}(2)$.
Moreover, each photon is detected with a lower probability due to
detectors efficiencies, fiber couplings, and filters.  Two Stokes photons can be produced
via two processes: two photons in one of the pulses and zero in the
second and one photon in each write pulse. For our experiment both
options are fine, as long as the creation of three photons will be
negligible. Since the process of creating N Stokes photons is a Poisson
process, we can calculate the probability of two and three photons
events using the probability of a one photon event. We note also that lower detection efficiency may cause three photons
to be detected as two, for example. In the case two photons are created the
probability to detect them is of course dependent upon the detection
efficiency $P_{detect}$ and will be in our case $P_{det}(2\text{,}2)=P_{\lambda}(2)\times P_{detect}^{2}$.
The probability of detecting only two photons while exciting $n$ atoms
will be $P_{det}(n\text{,}2)=\left(\begin{array}{c}
n\\
2
\end{array}\right)\times P_{\lambda}(n)\times P_{detect}^{2}\times(1-P_{detect})^{n-2}$. 
Dark counts also contribute to lower the fidelity by adding a false
detected photon. The probability for one dark count per pulse up to
first order will be $P_{dark}(1)=P_{\lambda}(0)\times P_{dc}+P_{\lambda}(1)\times(1-P_{detect})\times P_{dc}$,
where $P_{dc}$ is the probability for a dark count per pulse that
is related to the length of the pulse.

\begin{figure}[H]
\includegraphics[width=8.6cm]{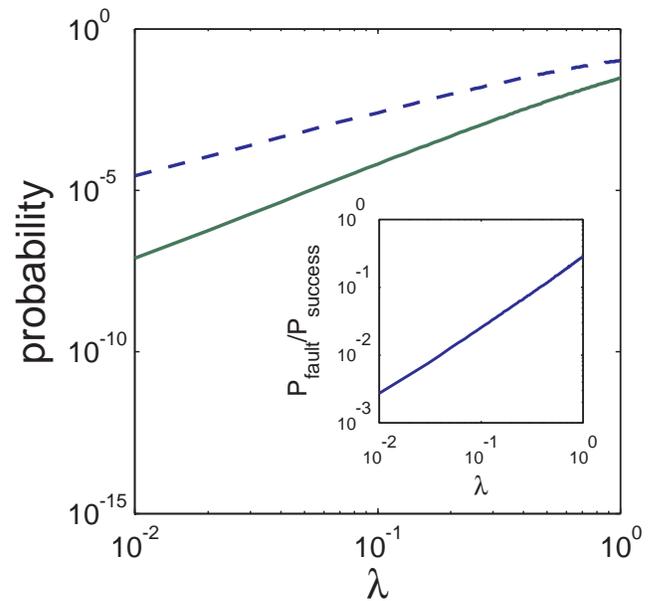}

\caption{\label{fig:succes_fault} Probability for a successful and false events
as a function of the Poisson parameter (blue - successful event, green
- false event). The calculation uses dark counts probability of $10^{-6}$
and detection efficiency of 75\%. Inset shows the ratio between the
events probabilities. }
\end{figure}

In order to quantify the total fidelity of the state created and the
rate of successful events we assume a post selected measurement where
we measure one AS photon after each read pulse. In this case
a successful event is regarded as an event where two Stokes photons
are created and detected with different polarization, thus the probability
of such an event is $P_{succes}=P_{det}(2,2)$. False events are all
the events with excitation number $n\neq2$ that lead to a detection
of two Stokes photons.  Figure \ref{fig:succes_fault} shows the probabilities
of false and successful events as a function of the Poisson parameter. Here we take
a detection efficiency of $P_{detect}=75\%$ and dark counts rate of 10 Hz \cite{Eisaman2011}, thus for a 100 ns pulse $P_{dc}=10^{-6}$.
The fidelity can be taken as the ratio between false and successful probabilities,
thus a fidelity of 95\% is achievable using a Poisson parameter $\lambda=0.2$.

\begin{figure}[H]
\includegraphics[width=8.6cm]{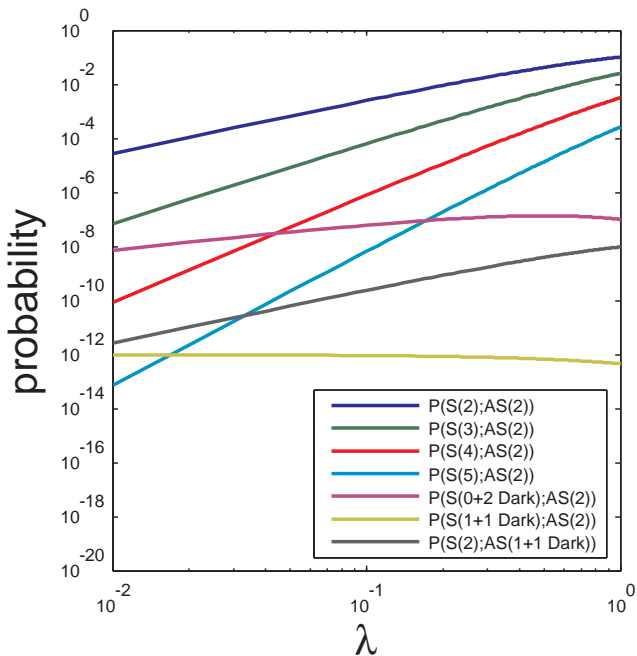}

\caption{\label{fig:probability graph}Probabilities of the main events that
can contribute to a post selected experiment when two Stokes photons
and two AS photons are detected. The notation P(S(n);AS(m))
represents the actual number of photons + dark counts that are created
in the experiment. P(S(2);AS(2)) is a successful event, while all others
contribute to false events. The sum of all these false events is shown
in Fig. \ref{fig:succes_fault}. }
\end{figure}

Figure \ref{fig:probability graph} presents the probabilities of
the main false events. The predominant false events in low $\lambda$
are dark counts in the detector while high $\lambda$ suffers mostly
from events related to false detection of higher excitation modes
due to the imperfect detection efficiency. It is important to notice
that there is a trade off between maximizing the rate of successful
events (larger $\lambda$) and minimizing the false detection of higher
events (smaller $\lambda$).

The possible rate of a two photon entanglement source (disregarding
the detection efficiency of the AS photon) can be estimated
by calculating the potential probability for a successful experiment
which is $\frac{1}{2}\times P_{det}(2,2,\lambda=0.2)\times (P_{B}(1))^2\approx10^{-3}$,
where the half is due to detections of $\left|+_{s}+_{s}\right\rangle /\left|-_{s}-_{s}\right\rangle $
and $P_{B}(1)=0.5$ is the optimal probability for one AS
emission according to Binomial distribution with a total of two excitations.
This ensures maximal rate for the read process. This probability will create an entangled pair with Fidelity of 95\%.  Assuming a repetition
rate of 10 MHz, bounded by pumping rate due to the natural lifetime of the atoms, the successful experiment rate will be $\sim$10 kHz. This
calculation takes into account the best up to date detectors, with minimal fiber coupling losses. A more conventional setup may have lower detection efficiencies of $\sim$30\%. This will lower the rate substantially to $\sim$100 Hz for the same fidelity. The
tremendous progress in the field of single photon detection  \cite{Eisaman2011} implies that even higher rates will be possible in the near future.

\section{Conclusions \label{sec:Conclusions}}

A scheme for the creation of a polarization entanglement between two
photons using atomic ensemble was presented. This scheme relies on
the fact that optical transitions between Zeeman sublevels in the
single photon regime may act as a polarization beam splitter. Taking experimental restrictions to consideration we estimate a fidelity that can reach up
to 95\% with entangled pair production rate of 10 kHz. Combined with the ability to store the photons
as a polariton in the ensemble this scheme has the potential to become
a useful source for quantum communication beyond current
available sources.

\section*{Acknowlegments \label{sec:Acknow}}
We acknowledge the support of ISF Bikura grant no. 1567/12. A. Retzker also acknowledge the support of carrier integration grant(CIG) no. 321798 IonQuanSense FP7-PEOPLE-2012-CIG.

\bibliographystyle{apsrev4-1}

\begin{thebibliography}{23}
\bibitem []{unity}%
    \BibitemOpen
    \bibinfo
    {note} {We assume here that the AS photon extraction probability is sufficiently bellow unity. If this is not 
    the case, corrections are needed for eq. \eqref{eq:6} due to a breaking of the symmetry between the extraction probabilities of the first and the second AS photons and the final state is no longer maximally entangled.}
     \BibitemShut {Stop} %
\end{thebibliography}
%

\end{document}